\magnification = \magstephalf
\baselineskip = 14truept

 at 12truept

\font\apj = cmcsc10
\nopagenumbers
\raggedbottom
\def\et{{\sl et al.\ }}

\def\EE#1 #2{$#1 \times 10^{#2}\, {\rm ergs\> s^{-1}}$}
\def\FF#1 #2{$#1 \times 10^{#2}\, {\rm ergs\> cm^{-2}\> s^{-1}}$}

\def\hi {\noindent \hangindent=2.5em}
\def\eg{{\sl e.g.,\ }}

\def\Msun{\ifmmode M_{\odot} \else $M_{\odot}$\fi}
\def\updna#1 #2 #3{$#1^{\scriptscriptstyle +#2}\!\!\!\!\!\!\!\!\!_{\scriptscriptstyle -#3}$}
\def\updnb#1 #2 #3{$#1^{\scriptscriptstyle +#2}\!\!\!\!\!\!\!\!\!\!\!_{\scriptscriptstyle -#3}$}
\def\updnex#1 #2 #3 #4{$#1^{\scriptscriptstyle +#2}\!\!\!\!\!\!\!\!\!\!\!_{\scriptscriptstyle -#3} \times 10^{#4}$}
\headline = {\tenrm\ifnum \pageno > 1 \centerline{-- \folio\ --}\else\hfil\fi}
\def\nts{\negthinspace}
\input epsf

\topglue 0.83truein

\centerline{THE 3--53 keV SPECTRUM OF THE QUASAR 1508+5714:~X-RAYS FROM $z = 4.3$}

\bigskip\bigskip
\centerline{\apj Edward C.~Moran$^1$ and David J.~Helfand$^2$}

\bigskip\bigskip\bigskip
\centerline{ABSTRACT}
\bigskip

We present a high-quality X-ray spectrum in the 3--53 keV rest-frame band of
the radio-loud quasar 1508+5714, by far the brightest known X-ray source at
$z > 4$.  A simple power-law model with an absorption column density equal to
the Galactic value in the direction of the source provides an excellent and
fully adequate fit to the data; the measured power-law photon index $\Gamma$
= \updnb {1.42} {0.13} {0.10}. Upper limits to Fe K$\alpha$ line emission and
Compton-reflection components are derived.  We offer evidence for both X-ray
and radio variability in this object and provide the first contemporaneous
radio spectrum ($\alpha = -0.25$). The data are all consistent with a picture
in which the emission from this source is dominated by a relativistically
beamed component in both the X-ray and radio bands.

\medskip\noindent
{\sl Subject headings:} galaxies:~active --- quasars:~individual (1508+5714)
--- X-rays:~galaxies

\vfootnote {$^1$}{Institute of Geophysics and Planetary Physics, Lawrence Livermore National Laboratory, \break\hskip2em L--413, Livermore, CA 94550}

\vfootnote {$^2$}{Department of Astronomy, Columbia University, 538 West 120th Street, New York, NY 10027}

\bigskip\bigskip\bigskip
\centerline{1.\ INTRODUCTION}
\bigskip

Surveys of the X-ray sky with the {\sl Einstein\/} and {\sl ROSAT\/}
observatories have revealed that, over a broad range of fluxes, quasars
are the most common extragalactic X-ray sources (Stocke \et 1991; Boyle
\et 1993). Thousands of predominantly low-redshift quasars have now been
observed with these instruments, providing a comprehensive picture of
their soft X-ray properties (\eg Ku, Helfand, \& Lucy 1980; Zamorani \et
1981; Avni \& Tananbaum 1986; Wilkes \& Elvis 1987; Wilkes \et 1994; Laor
\et 1994; Green \et 1995).  But investigation of the {\sl hard\/} X-ray
spectra of quasars has, until recently, only been possible for the
handful of nearby and exceptionally bright objects suitable for study with
nonimaging instruments, such as those on board {\sl EXOSAT\/} and
{\sl Ginga\/} (Lawson \et 1992; Williams \et 1992). Hence, comparatively
little is known about the characteristics of quasars above a few keV,
where most of their X-ray energy is emitted.

X-ray observations of high-redshift quasars provide access to their hard
X-ray spectra and, through comparison to low-redshift objects, the
opportunity to explore the evolution of their high-energy properties.  Recent
{\sl ROSAT\/} observations of $z \approx 3$ quasars in the 0.1--2.4 keV band
have served both functions, yielding spectra in the 0.5--10 keV rest frame
energy range for objects emitting when the universe was roughly one-quarter
its present age (Elvis \et 1994a; Bechtold \et 1994a; Pickering, Impey, \&
Foltz 1994).  Many of the same objects observed with {\sl ROSAT\/} have been
studied with the {\sl ASCA\/} satellite in order to examine their spectral
properties up to rest energies of $\sim$ 40 keV (Serlemitsos \et 1994; Elvis
\et 1994b; Siebert \et 1996; Cappi \et 1997).  Some preliminary conclusions
have been drawn about the X-ray spectral evolution of quasars ({\sl e.g.},
Bechtold \et 1994b), but to date just 25 objects with redshifts in excess
of 3 have been detected in the X-ray band, and spectral information is
available for only a fraction of these.  Thus, each new high-$z$ example
provides a valuable datum for quasar evolution studies.  In this
{\it Letter\/} we present the results of deep {\sl ASCA\/} observations
of the $z = 4.30$ quasar 1508+5714, the brightest known quasar in the 
high-redshift universe.

We discovered 1508+5714 and its X-ray emission as part of our follow-up of
unidentified radio-selected X-ray sources in the {\sl Einstein\/} Two-Sigma
Catalog (Moran \et 1996).  Despite the fact that we found the quasar to be
a relatively strong X-ray source (detected at the 6~$\sigma$ level in a
$\sim$~2800~s exposure), it had apparently escaped notice in all previous
analyses of the {\sl Einstein\/} IPC image.  Contemporaneous discovery of
1508+5714 was made by Hook \et (1995) in their sample of flat-spectrum radio
sources.  Spurred by the Hook \et report, Mathur \& Elvis (1995) reanalyzed
the {\sl Einstein\/} image containing the quasar and also found that it was
detected.  1508+5714 is nearly the most distant X-ray source known, second
only to RX~J1759.4+6638 at $z = 4.32$ (Henry \et 1994), which is more than 50
times fainter.  The only other $z > 4$ quasar detected at X-ray wavelengths
is 0000--263 ($z = 4.11$; Bechtold \et 1994a), which is ten times fainter than
1508+5714.  Thus, 1508+5714 currently provides the only opportunity to study
in detail quasar X-ray emission above a redshift of 4.  We report here a
measurement of its spectrum in the 0.5--10 keV {\sl ASCA\/} bandpass,
equivalent to the 3--53 keV band in the rest frame of the quasar.

\bigskip\medskip
\centerline{2.\ X-RAY AND RADIO OBSERVATIONS}
\bigskip

\centerline{2.1.~{\sl Broadband X-ray Observations}}
\medskip

1508+5714 was observed with the {\sl ASCA\/} satellite (Tanaka, Inoue, \&
Holt 1994) on two occasions, first on 2 March 1995 and then on 15 December
1995.  Data collected with both sets of instruments on board {\sl ASCA},
the Gas Imaging Spectrometers (GIS2 and GIS3) and the Solid-state Imaging
Spectrometers (SIS0 and SIS1), were filtered following the guidelines
described in The ABC Guide to {\sl ASCA\/} Data Reduction (Day \et 1995).
A total of 92.1~ks of good exposure was obtained with each of the GIS
detectors: 53.3~ks during the first observation and 38.8~ks during the
second.  Exposure times with the SIS instruments, which were operated in
1-CCD mode, totaled 83.7 and 83.0~ks for SIS0 and SIS1, respectively,
with 58\% of the SIS exposure acquired during the March observation.
1508+5714 was placed at slightly different positions on the detectors in
the two observations (the offset is $\sim$~$1'$), causing the source to be
vignetted by different amounts and making the extraction of spectral
information from the co-added images imprudent.  Therefore, for each
instrument we accumulated source and background spectra for the March and
December portions of the observation separately, and combined the spectra
afterwards using the FTOOLS software task ``addascaspec.''

We extracted source counts within a region $4'$ in radius centered on the
quasar in the GIS images and within a region $2'$\nts\nts.5 radius in
the SIS images.  Unfortunately, 1508+5714 lies just $3'$\nts\nts.5 to
the northeast of the nearby spiral galaxy NGC 5879 ($z = 0.0026$).  Although
the galaxy was not detected in the original {\sl Einstein\/} IPC image
(implying a 0.2--3.5 keV flux of $<$ 17\% that of the quasar), it could
conceivably contaminate the spectrum of 1508+5714 in the considerably deeper
{\sl ASCA\/} observation.  However, as the combined SIS0 + SIS1 image from
the March observation indicates (see Fig.~1), NGC 5879 is unlikely to
contribute to the SIS spectrum of the quasar.  The larger extraction region
needed for the GIS spectrum does include the position of
the galaxy, but our separate fits to the SIS and GIS spectra (\S~3.1) yield
very similar results, suggesting that NGC 5879 does not significantly
contaminate the GIS spectrum of 1508+5714 either.

To measure the GIS background, we collected counts within source-free regions
at the same distance off-axis as the quasar and with twice the area of the
source region. We estimated the background in the SIS spectra by extracting
counts from the entire chip, omitting a region $4'$ in radius around the
source.  The total number of background-subtracted counts obtained in the
SIS0, SIS1, GIS2, and GIS3 spectra of 1508+5714 in the 0.5--10 keV band are
1294, 1104, 808, and 1097, respectively.  To improve the signal-to-noise ratios
of the spectra for model fitting, we used the ``addascaspec'' program once more
to combine the SIS0 spectrum with the SIS1 spectrum and the GIS2 spectrum with
the GIS3 spectrum. These spectra, which are referred to as the SIS and GIS
spectra below, were binned to have at least 100 counts (source plus background)
per channel.

{\vskip 0.4truein
\hskip 0.8truein
\epsfxsize=4.25truein
\epsffile[86 216 531 630]{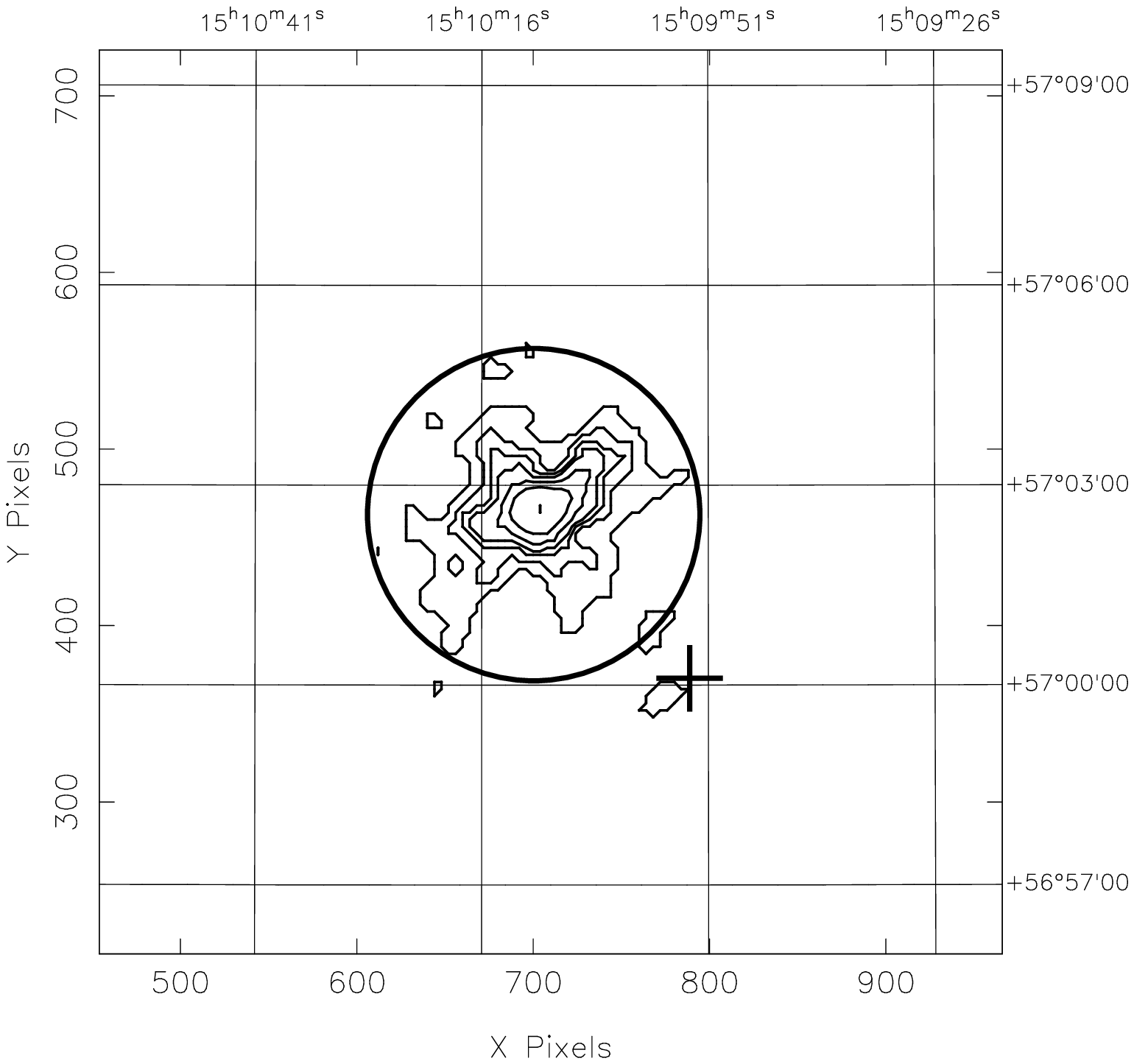}

\bigskip
{\narrower {\apj Fig.}~1.---Contour plot of the combined {\sl ASCA\/} SIS0 and
SIS1 images of 1508+5714 from the March 1995 observation.  The image has been
smoothed with a 24$''$ FWHM Gaussian, and contours at the 3, 4, 5, 6, 8, 11,
and 21 $\sigma$ level are plotted.  Owing to spacecraft aspect uncertainties,
the peak of the X-ray emission is located 45$''$ east of the optical position
of the quasar, so we have shifted the
position of the nearby spiral galaxy NGC 5879 (Dressel \& Condon 1976),
indicated by the cross, accordingly.  Although NGC 5879 may have been
marginally detected in this deep {\sl ASCA\/} observation, the galaxy lies
outside the circular region used to extract source counts and is unlikely
to contaminate the SIS spectrum of the quasar.\par}}

\bigskip
\centerline{2.2.~{\sl Radio Continuum Observations}}
\medskip

Over the past two decades, 1508+5714 has been detected at a variety of radio
frequencies, including 365 MHz ($S = 191$ mJy; Douglas \et 1996), 1.4 GHz
($S = 149$ mJy; White \& Becker 1992, and $S = 202$ mJy; Condon \et 1997),
4.9 GHz ($S = 279$ mJy; Becker, White, \& Edwards 1991), and 8.4 GHz ($S =
153$ mJy; Patnaik \et 1992).  Given the difference between the two 1.4~GHz
measurements, it appears that the quasar is a variable radio
source.  To investigate its radio properties further, we observed 1508+5714
with the Very Large Array (VLA) in the A configuration on 13 July 1995.  The
source was observed for 5 minutes each at 1.46 GHz and 8.41 GHz.  The quasar
is unresolved at both frequencies, with measured flux densities of 234 mJy
at 1.46 GHz and 152 mJy at 8.41 GHz.  These observations confirm the radio
variability of 1508+5714 and provide the first contemporaneous measurement
of its radio spectrum, which is moderately flat: assuming $S_{\nu} \sim
\nu^{\alpha}$, $\alpha = -0.25$.

\bigskip\medskip
\centerline{3.\ RESULTS}

\bigskip
\centerline{3.1.~{\sl The X-ray Spectrum of 1508+5714}}
\medskip

As Table 1 indicates, a simple model consisting of a single absorbed power
law provides an excellent fit to the {\sl ASCA\/} spectrum of 1508+5714.  The
results obtained for separate and simultaneous fits to the SIS and GIS
spectra are very similar.  For the simultaneous fit, the derived photon index
and column density, assuming the absorber is at $z = 0$, are $\Gamma$ = \updnb
{1.42} {0.13} {0.10} and $N_{\rm H}$ = \updna {1.4} {5.4} {1.4} $\times
10^{20}$ cm$^{-2}$.  (All errors listed are at the 90\% confidence level for
two interesting parameters, unless otherwise noted.)  This column density
is consistent with the Galactic column in the direction of 1508+5714 of
$1.6 \times 10^{20}$ cm$^{-2}$ (Stark \et 1992).  Although an additional
component of absorption at the redshift of the quasar is not required, we
can derive an upper limit to the column density of such a component:
$1.3 \times 10^{22}$ cm$^{-2}$.  In light of recent
evidence that excess absorption is common in the spectra of radio-loud
high-$z$ quasars (Cappi \et 1997), the lack of excess absorption in the

\bigskip\bigskip
{\hsize = 8.5truein\parindent 1.5em
\hskip 0.15truein\vbox{
\halign{#\hfil\tabskip 2em &  \hfil#\hfil & \hfil#\hfil & \hfil#\hfil & \hfil#\hfil & \hfil#\hfil \tabskip 0pt \cr
\multispan{6}\hfil\apj TABLE 1\hfil\cr\noalign{\vskip 4pt}
\multispan{6}\hfil\apj Power-Law Fits to the {\sl ASCA\/} Spectra of 1508+5714\hfil\cr
\noalign{\vskip 1em\hrule\vskip 2pt\hrule\vskip 1em}
 & Energy Range& & $N_{\rm H}$& \cr
 Instruments& (keV)& $\Gamma$& ($\times 10^{20}$ cm$^{-2}$)& $A^{\rm a}$& $\chi^2$ (d.o.f)\cr
\noalign{\vskip 1em\hrule\vskip 1em}
SIS + GIS& 0.5--10& \updnb {1.42} {0.13} {0.10}& \updna {1.4} {5.4} {1.4}& 1.14& 53.9 (58)\cr
SIS      & 0.5--10& \updnb {1.45} {0.17} {0.12}& \updna {1.2} {6.1} {1.2}& 1.12& 27.8 (27)\cr
GIS$^{\rm b}$ & 0.7--9& \updnb {1.41} {0.10} {0.10}& 1.2$^{\rm c}$~~ & 1.17& 22.6 (29)\cr
SIS + GIS&  0.5--2& 1.43~~~~        & 1.4$^{\rm c}$~~ & 1.14& 39.0 (34)\cr
SIS + GIS&   2--10& 1.42~~~~        & 1.4$^{\rm c}$~~ & 1.13& 27.0 (31)\cr
\cr
\noalign{\hrule}
\noalign{\vbox{\hsize=6.25truein

\vskip 1em
$^{\rm a}$ Power-law normalization at 1 keV, in units of $10^{-4}$ photons cm$^{-2}$ s$^{-1}$ keV$^{-1}$.

$^{\rm b}$ For this fit, errors on $\Gamma$ are 90\% confidence for one interesting parameter.

$^{\rm c}$ Parameter fixed.
}}}}\par}

\noindent
spectrum of 1508+5714 is somewhat surprising.  Assuming there is nothing
special about the universe in the direction of 1508+5714, the existence
of a comparatively clear line of sight out to $z = 4.3$ supports the
prevailing interpretation that excess absorption in the spectra of other
radio-loud quasars is intrinsic to those objects (Cappi \et 1997).
The observed flux of 1508+5714 in the 0.5--10 keV range is \FF {9.8} {-13}.
For $H_0 = 50$ km s$^{-1}$ Mpc$^{-1}$ and $q_0$ = 0, the implied isotropic
luminosity of the quasar is \EE {2.2} {47} in the (rest frame) 2--10 keV
band, and \EE {7.7} {47} in the 2--50 keV band. The {\sl ASCA\/} spectrum
of 1508+5714 and the power-law fit described above are displayed in Figure~2.

{\vskip -0.8truein\hskip 0.6truein
\epsfxsize=4.5truein
\epsffile{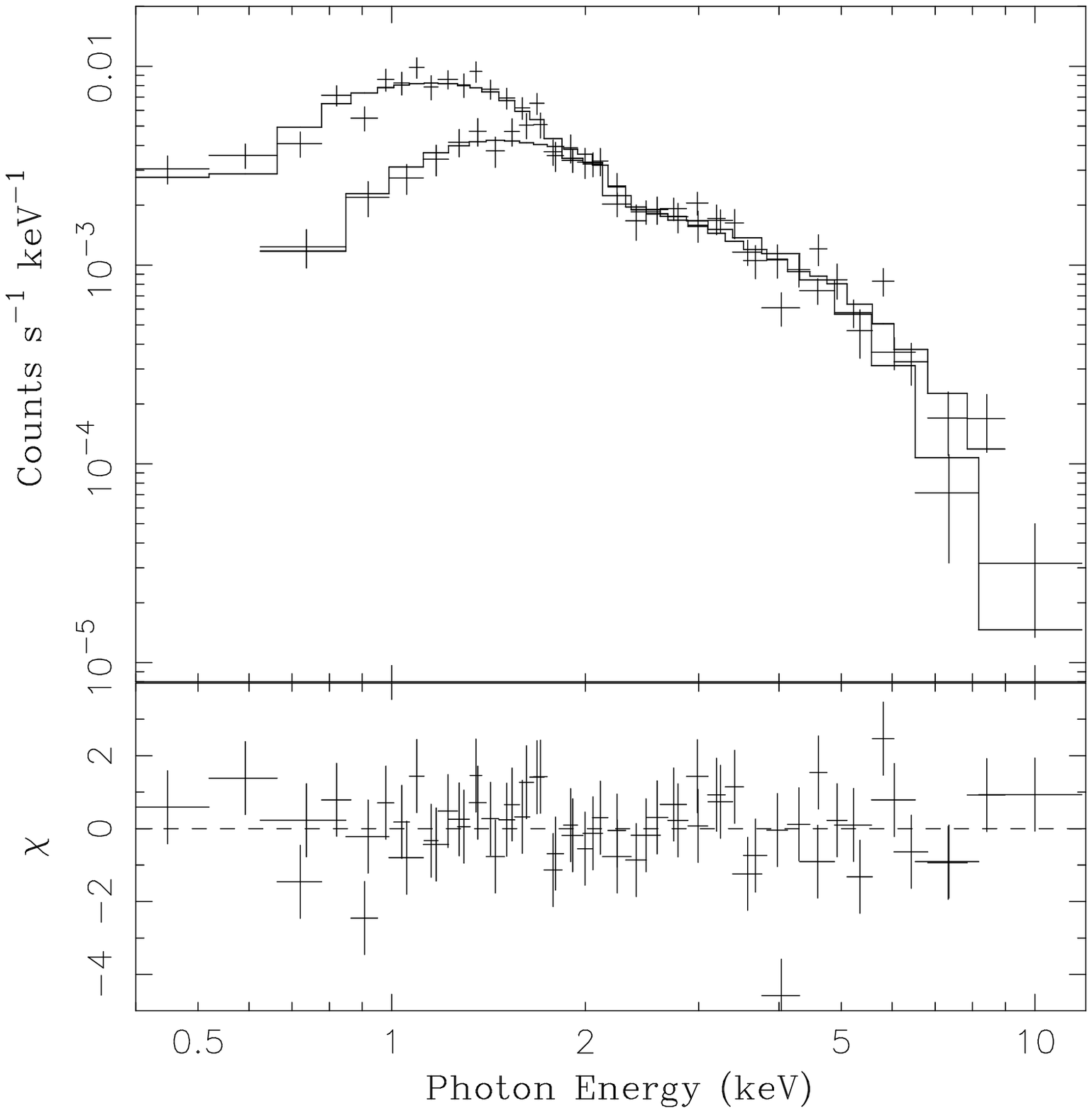}

\vskip -1.0truein
{\narrower 
{\apj Fig.}~2.---The observed {\sl ASCA\/} SIS and GIS spectra of 1508+5714
and the best fitting power-law model.\par}}

\bigskip\smallskip
The derived power-law photon index $\Gamma = 1.4$ is consistent with the
mean spectrum of core-dominated radio-loud quasars (Worrall \& Wilkes 1990).
Additional spectral components, such as a 6.4 keV Fe fluorescence line
or a Compton-reflection hump, which would flatten the X-ray
spectrum above $\sim$~10 keV and peak at a rest energy of 20--30 keV
(Lightman \& White 1988), are not required to fit the spectrum of 1508+5714.
As a first attempt to investigate whether or not the continuum of 1508+5714
possesses any complex features, we measured the power-law index of the
spectrum above and below 2 keV, which corresponds to $\sim$~10 keV in the
rest frame of the quasar.  As indicated in Table 1, the slope of the quasar's
spectrum is remarkably consistent over its entire range.  When the spectrum
is fitted with a Compton-reflection model, an upper limit of 0.5 is obtained
for the fractional solid angle $\Omega/2\pi$ subtended by the reflector,
assumed to be a disk.  For reference, values of $\Omega/2\pi$ in the range
0.5--1 for are found for low-$z$ Seyfert 1 galaxies exhibiting reflection
components in their spectra (Nandra \& Pounds 1994).  The 90\% confidence
upper limit to the equivalent width of an intrinsically narrow Fe K$\alpha$
emission line is 31 eV at an observed energy of 1.21 keV, which translates
to 167 eV in the quasar frame.

\bigskip
\centerline{3.2.~{\sl X-ray Variability}}
\medskip

The two {\sl ASCA\/} observations of 1508+5714, separated by 9 months,
provide the opportunity to search for X-ray variability on a timescale of
54 days in the frame of this object.  Using the ``addascaspec'' task, we
combined the SIS0 spectrum with the SIS1 spectrum and the GIS2 spectrum with
the GIS3 spectrum for both the March and December segments of the observation.
To test for variability, we applied a power-law model to the spectra from
the two epochs separately, fixing the absorption column density at the
Galactic value.  We find that the spectrum and intensity of 1508+5714 varied
significantly between the two observations: a $\Gamma$ = \updnb {1.55} {0.09}
{0.08} power law with a 1~keV normalization of \updnex {1.39} {0.10} {0.10}
{-4} photons cm$^{-2}$ s$^{-1}$ keV$^{-1}$ fits the March spectrum, whereas
a $\Gamma$ = \updnb {1.25} {0.14} {0.15} power law with a normalization of
\updnex {0.80} {0.09} {0.12} {-4} photons cm$^{-2}$ s$^{-1}$ keV$^{-1}$ fits
the December spectrum.  The implied decrease in the 0.5--10 keV flux of
1508+5714 over the span between the two observations is 15\%.
Since we extracted the March and December spectra separately and determined
the correct effective area for each, we are certain that the variability is
not a consequence of the vignetting differences associated with the different
off-axis positions of the quasar in the two exposures.

The X-ray variability of 1508+5714 could, in principle, allow us to place
limits on the mass of the central black hole, which would be of interest for
an object at such a high redshift.  However, there are several lines of
evidence which point to beaming as the origin of variability in 1508+5714.
Not only is the quasar's flat-spectrum radio
source variable, it is unresolved at a resolution of $\sim$~5 mas in a
5 GHz VLBI observation (Frey \et 1997).  The compactness and high flux density
(286 mJy) of the VLBI source, in addition to the radio variability, suggest
that the radio emission from 1508+5714 is highly beamed.

As discussed by Mathur \& Elvis (1995), 1508+5714 appears to be considerably
overbright in X-rays given its optical luminosity, implying that a significant
fraction of its X-ray emission is beamed as well.  By extrapolating the mean
unabsorbed {\sl ASCA\/} spectrum of 1508+5714 to the energy corresponding to
2 keV rest and our optical spectrum (Moran \et 1996) to the wavelength
corresponding to 2500~\AA\ rest (assuming $f_{\rm opt} \sim \nu^{0.5}$), we
obtain a value of 0.93 for the two-point optical--to--X-ray spectral slope
$\alpha_{\rm ox}$ (Tananbaum \et 1979), which confirms the extreme X-ray
brightness of 1508+5714 (see Wilkes \et 1994).  Thus, the X-ray luminosities
of 1508+5714 given above in \S~3.1 are not likely to be isotropic values.  A
viewing geometry close to the axis of the radio jet might also explain why
excess absorption is not observed in the X-ray spectrum of 1508+5714.

\bigskip
\centerline{3.3.~{\sl Implications for the X-ray Background}}
\medskip

The spectrum of the cosmic X-ray background (XRB) is well fitted by a
$\Gamma = 1.4$ power law in the {\sl ASCA\/} band (Gendreau \et 1995), and by
a $kT = 40$ keV thermal bremsstrahlung model in the 3--50 keV range (Marshall
\et 1980).  It is now certain that the XRB arises from the integrated emission
of discrete sources (Mather \et 1990; Wright \et 1994), even though a class of
objects with an average spectrum similar to that of the XRB has yet to be
identified; moderate-redshift quasars and Seyfert galaxies, the most numerous
extragalactic sources at bright X-ray fluxes, have collective spectra which
are generally far too steep to explain the background spectrum (Fabian \&
Barcons 1992).  Interestingly, the application of redshifted thermal models
to the {\sl ASCA\/} spectra of $z \approx 3$ radio-loud quasars have revealed
rest-frame temperatures in the 34--45 keV range (Serlemitsos \et 1994; Elvis
\et 1994b; Siebert \et 1996), suggesting to these authors that AGNs may
produce the XRB after all. The similarity between the spectrum of 1508+5714
($\Gamma = 1.4$) and the spectrum of the XRB might be cited to support this
hypothesis. But in addition to the obvious difficulties quasars have satisfying
the integrated luminosity and areal surface density requirements imposed on
the XRB-producing class of sources,
a redshifted bremsstrahlung model for 1508+5714 (with $N_{\rm H}$ fixed
at the Galactic value) yields a rest-frame temperature of 93 keV, with a
90\% confidence range of 69--133 keV.  A temperature of 40 keV is ruled
out at $>$~99\% confidence.

\bigskip\medskip
\centerline{4.\ SUMMARY}

\bigskip
Since sensitivity improvements of an order of magnitude over the best existing
detectors (\eg {\sl XTE}) will be required to obtain high-quality spectra of
a substantial sample of low-redshift AGNs in the 10--50 keV band, high-redshift
quasars currently offer the only window available in this important energy
regime.  We have measured the spectrum of the brightest X-ray source known at
$z > 4$, 1508+5714, and find that it is well-described by a simple power law
with a photon index of 1.4 and no requirement for absorption in excess of the
Galactic neutral hydrogen column density.
Our measurement of a flat radio spectral index and the evidence we adduce for
both radio and X-ray variability suggest that relativistically beamed emission
contributes significantly to this quasar's energy budget; the limits we derive
for Fe fluorescence and Compton-reflection components are understandable in
this context, since the beamed radiation is likely to dominate the isotropic
X-ray flux from the quasar.

Our {\sl ASCA\/} observation of 1508+5714 broadens considerably the range of
redshifts over which the X-ray properties of quasars have been determined.
However, the properties of 1508+5714 are entirely consistent with those of
other radio-loud X-ray quasars (see Fig.~3 of Cappi \et 1997), which
supports preliminary conclusions that radio-loud quasars exhibit no spectral
evolution with redshift or luminosity (Cappi \et 1997).  Of course, additional
examples of X-ray--bright high-redshift quasars are needed to address this
question fully.  The advent of {\sl AXAF\/} and {\sl XMM\/} in the next few
years will allow the extension of studies exploring the hard X-ray spectra
of AGNs to many more such denizens of the high-$z$ universe.

\bigskip
We are grateful to Leonid Gurvits for information about the VLBI observations
of 1508+5714, to Sally Laurent-Muehleisen for illuminating discussions about
relativistic beaming in AGNs, and to Bob Becker for assistance with the VLA
data.  This work has been supported at Columbia by a grant from NASA under
the {\sl ASCA\/} Guest Investigator Program (NAG5--2556) and at LLNL by the
US Deprtment of Energy under contract W-7405-ENG-48.  This is contribution
number 629 of the Columbia Astrophysics Laboratory.

\bigskip\medskip
\centerline{\apj REFERENCES}
\bigskip

\hi Avni, Y., \& Tananbaum, H.\ 1986, ApJ, 305, 83

\hi Bechtold, J., et al.\ 1994a, AJ, 108, 374

\hi ---------.~1994b, AJ, 108, 759

\hi Becker, R.\ H., White, R.\ L., \& Edwards, A.\ L.\ 1991, ApJS, 75, 1

\hi Boyle, B.\ J., Griffiths, R.\ E., Shanks, T., Stewart, G.\ C., \& Georgantopoulos, I.\ 1993, MNRAS, 260, 49

\hi Cappi, M., Matsuoka, M., Comastri, A., Brinkmann, W., Elvis, M., Palumbo, G.\ G.\ C., \& Vignali, C.\ 1997, ApJ, 478, 492

\hi Condon, J.\ J., Cotton, W.\ D., Greisen, E.\ W., Yin, Q.\ F., Perley, R.\ A., Taylor, G.\ B., \& Broderick, J.\ J.\ 1997, preprint

\hi Day, C., Arnaud, K., Ebisawa, K., Gotthelf, E., Ingham, J., Mukai, K., \& White, N.\ 1995, The ABC Guide to {\sl ASCA\/} Data Reduction (Greenbelt:~NASA/GSFC)

\hi Douglas, J.\ N., Bash, F.\ N., Bozyan, F.\ A., Torrence, G.\ W., \& Wolfe, C.\ 1996, AJ, 111, 1945

\hi Dressel, L.\ L., \& Condon, J.\ J.\ 1976, ApJS, 31, 187

\hi Elvis, M., Fiore, F., Wilkes, B., McDowell, J., \& Bechtold, J.\ 1994a, ApJ, 422, 60

\hi Elvis, M., Matsuoka, M., Sieminginowska, A., Fiore, F., Mihara, T., \& Brinkmann, W.\ 1994b, ApJ, 436, L55

\hi Fabian, A.\ C., \& Barcons, X.\ 1992, ARA\&A, 30, 429

\hi Frey, S., Gurvits, L.\ I., Kellermann, K.\ I., Schilizzi, R.\ T., \& Pauliny-Toth, I.\ I.\ K.\ 1997, A\&A, in press

\hi Gendreau, K.\ C., et al.\ 1995, PASJ, 47, L5

\hi Green, P.\ J., et al.\ 1995, ApJ, 450, 51

\hi Henry, J.\ P., et al.\ 1994, AJ, 107, 1270

\hi Hook, I.\ M., McMahon, R.\ G., Patnaik, A.\ R., Browne, I.\ W.\ A., Wilkinson, P.\ N., Irwin, M.\ J., \& Hazard, C.\ 1995, MNRAS, 273, L63

\hi Ku, W.\ H-M., Helfand, D.\ J., \& Lucy., L.\ B.\ 1980, Nature, 288, 323

\hi Lawson, A.\ J., Turner, M.\ J.\ L., Williams, O.\ R., Stewart, G.\ C., \& Saxton, R.\ D.\ 1992, MNRAS, 259, 743

\hi Lightman, A.\ P., \& White, T.\ R.\ 1988, ApJ, 335, 57

\hi Marshall, F.\ E., et al.\ 1980, ApJ, 235, 4

\hi Mather, J.\ C.\ et al.\ 1990, ApJ, 354, L37

\hi Mathur, S., \& Elvis, M.\ 1995, AJ, 110, 1551

\hi Moran, E.\ C., Helfand, D.\ J., Becker, R.\ H., \& White, R.\ L.\ 1996a, ApJ, 461, 127

\hi Nandra, K., \& Pounds, K.\ A.\ 1994, MNRAS, 268, 405

\hi Patnaik, A.\ R., Browne, I.\ W.\ A., Wilkinson, P.\ N., \& Wrobel, J.\ M.\ 1992, MNRAS, 254, 655

\hi Pickering, T.\ E., Impey, C.\ D., \& Foltz, C.\ B.\ 1994, AJ, 108, 1542

\hi Serlemitsos, P., Yaqoob, T., Ricker, G., Woo, J., Kunieda, H., Terashima, Y., \& Iwasawa, K.\ 1994, PASJ, 46, L43

\hi Siebert, J., Matsuoka, M., Brinkmann, W., Cappi, M., Mihara, T., \& Takahashi, T.\ 1996, A\&A, 307, 8

\hi Stark, A.\ A., Gammie, C.\ F., Wilson, R.\ W., Bally, J., Linke, R.\ A., Heiles, C., \& Hurwitz, M.\ 1992, ApJS, 79, 77

\hi Stocke, J.\ T., et al.\ 1991, ApJS, 76, 813

\hi Tanaka, Y., Inoue, H., \& Holt, S.\ S.\ 1994, PASJ, 46, L37

\hi Tananbaum, H., et al.\ 1979, ApJ, 234, L9

\hi White, R.\ L., \& Becker, R.\ H.\ 1992, ApJS, 79, 331

\hi Wilkes, B.\ J.,\& Elvis, M.\ 1987, ApJ, 323, 243

\hi Wilkes, B.\ J., Tananbaum, H., Worrall, D.\ M., Avni, Y., Oey, M.\ S., \& Flanagan, J.\ 1994, ApJS, 92, 53

\hi Williams, O.\ R., et al.\ 1992, ApJ, 389, 157

\hi Worrall, D.\ M., \& Wilkes, B.\ J.\ 1990, ApJ, 360, 396

\hi Wright, E.\ L., et al.\ 1994, ApJ, 420, 450

\hi Zamorani, G., et al.\ 1981, ApJ, 245, 357

\bye